# A Knowledge-based Automated Debugger in Learning System

*Abdullah Mohd Zin, Syed Ahmad **Aljunid**\*, Zarina Shukur, **Mohd Jan Nordin***
*Universiti Kebangsaan Malaysia (UKM)*
*43600 Bangi, Selangor, Malaysia*
*(amz@sun1.ftsm.ukm.my)*

7 July 2000

*Abstract*

*Currently, programming instructors continually face the problem of helping to debug students' programs. Although there currently exist a number of debuggers and debugging tools in various platforms, most of these projects or products are crafted through the needs of software maintenance, and not through the perspective of teaching of programming. Moreover, most debuggers are too general, meant for experts as well as not user-friendly. We propose a new knowledge-based automated debugger to be used as a user-friendly tool by the students to self-debug their own programs. Stereotyped code (cliché) and bugs cliché will be stored as library of plans in the knowledge-base. Recognition of correct code or bugs is based on pattern matching and constraint satisfaction. Given a syntax error-free program and its specification, this debugger called **Adil** (**A**utomated **D**ebugger **i**n **L**earning system) will be able locate, pinpoint and explain logical errors of programs. If there are no errors, it will be able to explain the meaning of the program. Adil is based on the design of the **Conceiver**, an automated program understanding system developed at Universiti Kebangsaan Malaysia.*

1.     **The Problem**

Most students taking their first year course in programming have to face an arduous task of laboriously debugging their own programs. Although most compilers currently have built-in integrated debugging environment, these tools are not exactly user-friendly. Moreover, using watches, breakpoints, stepping, and other debugging aids require certain debugging skills too. Most of these tools are meant for experts. These novices usually turn to their seniors or their instructors to solve their dilemma. As the instructor-student ratio increases, the instructors are stuck in dealing with these trivial but time-consuming task.

In Malaysian universities, most of our first year students are able to write syntax-error free programs, albeit after some compiling iterations for some. Their main plight is in debugging the logical errors. As they have yet to acquire the basic debugging skills, they cannot exploit the full potential of the debugging utilities of the compiler.

Our solution for both these novices as well as for the instructors is to develop a knowledge-based automated debugger. This system called **Adil** (**A**utomated **D**ebugger **i**n **L**earning system), will assist the students in acquiring the basic debugging skills by guiding them in debugging their programs. Adil will localize, pinpoint and explain the bugs to them. In addition, Adil will aid them in the program understanding process which will be practically realized through the debugging sessions with Adil. For the instructors, Adil will also help to alleviate most of the manual and time-consuming debugging task currently being done by them by functioning as an intelligent debugging assistant.

---

\* is also attached to Universiti Teknologi MARA (UiTM), 40450 Shah Alam, Selangor, Malaysia (aljunid@tmsk.itm.edu.my)

The Adil prototype currently being developed will debug a subset of the C programming language. The source code will be parsed into an annotated flow graph. The annotated flow graph with data and control flow information will be fed into the inference engine and checked with the library of plans in the knowledge base for the unification of the plans and the code. The inference engine will either understand the error-free program, or locate the logical bug(s), pinpoint and explain them, if they exist, in the code as output of the debugging system.

## 2. Program Understanding and Debugging

Generally, program understanding is the process of acquiring knowledge about computer program. Specifically, *program understanding* is the process of recognizing program plans and extracting design goals in the source code [QYW98, KN94, Will92]. *Program plans* are abstract representations of the clichés or particular code patterns.

Understanding even a small program is a complex task that requires both knowledge and analysis. Indeed, the pattern matching algorithm between plans (represented by schemas, knowledge constraints or plan libraries) and programs (represented by the actual program, annotated abstract syntax trees or flow graphs) has been proven to be NP-Hard [WY96] as well as exponential in the worst case [Will92].

*Debugging* is the task of identifying and correcting faults in code. Once a program failure has been found, we must acquire an understanding of the program so as to localize the program fault and thus identify the program failure. Thus, the twin goals of debugging is to simultaneously localize the fault area of the code and develop an understanding of the program so that adequate correction can be applied [FR99]. However, debugging is an awesome labor-intensive and time-consuming activity. Hence, it is critical to develop an automatic debugger for this purpose.

## 3. Related Works

### 3.1 Classification of Automated Debugging Strategies

Ducassé proposed a classification of debugging strategies based on the type of actions to be taken. These are verification with respect to specification, checking with respect to language knowledge and filtering with respect to symptom [Duca93]. *Verification* is based on some formal specification of the intended program. It is the only strategy among the three that aims at fully automating the debugging process. *Checking* with respect to language knowledge systematically parses programs and searches for language dependent errors. Checking contains intrinsic limitations and should be used as as a complementary debugging strategy only. *Filtering* with respect to a symptom does not suspect any code but is merely aim at cutting the amount of code to be searched. As it is focus on locating the root of the error symptoms, it is more accurate for diagnosis purposes than the previous two strategies.

Meanwhile, Shahmehri's classification of debugging strategies is generally based on the approach used, i.e. static or dynamic [Shah91]. Specifically, Shahmehri acknowledged that these strategies are influenced by the type of knowledge given to the debugger as input as well as the type of tasks to be performed by the debugger. Common sources of debugger knowledge are actual program, actual program behavior, the intended program, the intended program behavior, the programming language, common types of bugs and the expertise level of the user. The tasks to be performed by the debugger are test generation, bug detection, bug localization, bug explanation and bug correction.

Automated debuggers use a combination of the above knowledge about the actual versus the intended programs to perform the debugging process. Common combinations are (a) actual vs. intended implementation (b) actual vs. intended input/output (c) actual vs. intended program behavior. She summarized that *static debugging approach* use any or all of the knowledge type mentioned above except for intended program behavior while *dynamic debugging approach* use the knowledge of the intended program behavior.

## 3.2 Automated Debugging Projects for Tutoring Systems

Automatic debugging systems practically consists of two main categories, tutoring systems and diagnosis systems. Generally, tutoring systems are based on static debugging strategy while diagnosis systems are based on dynamic debugging strategy. We are primarily interested in tutoring systems only in this paper.

While there have been a great deal of research done for diagnosis sytems, research in tutoring systems is still developing slowly. In fact, most of the surveyed automated tutoring systems for debugging were reported almost a decade ago by Ruth [Ruth76], Lukey [Luke80], Adam and Laurent [AL80], Johnson and Soloway [JS85], Murray [Murr88], Wertz [Wert87], Looi [Looi88], Harandi and Ning [HN88], and Allemang [Alle91]. An exception is the Bradman transparency debugger by Smith and Webb [SW95]. For comparison, we will utilize the criteria as spelt out above in 3.1.

Among the conclusions we can draw from Table 1 and Table 2 below are:
1. The target programming language is either a declarative or an imperative language.
2. Most systems locate and correct bugs but do not explain them.
3. All these systems implement the static debugging strategy except Bradman, which does language checking only.
4. Actual vs intended implementation is the combination of knowledge favored by most systems.
5. Most systems are programming language dependent.
6. Automatic knowledge base construction is only done by Pat.
7. All systems except for Phenarete and Bradman utilized the verification strategy.

## 4. Conceiver System

Conceiver is a knowledge-based program understanding system using constraint satisfaction developed by Universiti Kebangsaan Malaysia [Al-Om99]. Its framework is largely influenced by the Recognizer project by Rich and Wills [RW90]. It uses the bottom-up approach for automatic understanding of programs as well as for constructing the plan base while it applies the top-down approach for analyzing stereotype algorithms. The automatic understanding process is further broken into two steps; the bottom-up unification of code against the plans, and testing the successfully unified code against the constraints provided inside the unified plans.

The Conceiver (Fig. 1) consists of four main components: the infrastructure tools, the knowledge base, the understanding inference engine and the document generator. The prototype is implemented for Pascal language under the Wintel platform.

## 5. Adil System

The objectives of the Adil system are as follows:
(a) To improvise the Conceiver program understanding system, particularly the system's transformer as well as extending the plan formalism to include bugs cliché.
(b) To design a new knowledge-based automatic debugger which can understand correct code and/or debug logical errors of a given syntax error-free program based on a specific problem.
(c) To intelligently assist the students in debugging their programs by localizing, pinpointing and explaining the bugs.
(d) To intelligently assist the instructors by automatically acquiring plans and constructing the knowledge base.

Adil will use a subset of C, an imperative language as the target programming language. We will employ the static debugging strategy, in particular, we will utilize the actual vs intended implementation knowledge, as well as incorporating the programming language, programming expertise and common bug types knowledge. Although verification will be the backbone strategy, filtering will be added as a complementary strategy.

| Name | Target Programming Language | Developed by (University /Institute) | Principal Researcher(s) | Debugging task | Year reported | Knowledge combination utilized | V | C | F |
|---|---|---|---|---|---|---|---|---|---|
| **IPA** | Lisp/PL1 | MIT | G.R. Ruth | - | 1976 | static: actual vs intended implementation | x | - | - |
| **Pudsy** | Pascal | Sussex | F.J. Lukey | localize and correct bugs | 1980 | static: actual implementation vs intended i/o | x | x | - |
| **Laura** | FORTRAN | Caen | A.Adam, J.-P. Laurent | localize and explain bugs | 1980 | static: actual vs intended implementation | x | - | - |
| **Proust** | Pascal | Yale | W.L. Johnson, E. Soloway | localize, explain and correct bugs | 1985 | static: actual vs intended implementation | x | x | x |
| **Talus** | Lisp | Texas | W.R. Murray | localize and correct bugs | 1986 | static: actual vs intended implementation + intended i/o | x | - | - |
| **Phenarete** | Lisp | Paris VIII | H. Wertz | - | 1987 | static: actual vs intended implementation | - | x | - |
| **Apropos** | Prolog | Edinburgh | C-K. Looi | - | 1988 | static: actual vs intended implementation | x | x | - |
| **Pat** | Pascal-like | Illinois | M. Harandi, J. Ning | understand code, localize bugs | 1988 | static: actual vs intended implementation | x |  | - |
| **DUDU** | - | Ohio State | D. Allemang | identify correct code, explain and correct bugs | 1991 | static: actual vs intended implementation | x | - | - |
| **Bradman** | C | Deakin | P. Smith, G. Webb | localize & explain bugs (run-time) | 1995 | Dynamic: actual vs. language run-time | - | x | - |
| **Adil** | C | Universiti Kebangsaan Malaysia | S.A. Aljunid, A. Mohd Zin | understand code, localize & explain bugs | 2000 | static: actual vs intended implementation | x | - | x |

**Table 1** : A comparison of automated debuggers as tutoring systems

| Name | Other Knowledge Utilized* | Programming Language Independence Debugging System | Automatic Knowledge base construction | Internal Representation / Other Information |
|---|---|---|---|---|
| **IPA** | pl + cbt |  | No | plans |
| **Pudsy** | pl + pe | No | No | language dependent model schema |
| **Laura** | pl + pe + ad | Yes | No | language independent model graph-based |
| **Proust** | pl + pe + cbt | No | No | language dependent model |
| **Talus** | pl | No | No | language dependent model E-frames |
| **Phenarete** | pl | NA | NA | plans specialists (for syntax and semantics) |
| **Apropos** | pl + cbt | NA | NA | - |
| **Pat** | pe | Yes | Yes | language independent model events (object-oriented), plan-based, heuristics-based recognition |
| **DUDU** | pe | No | No | language dependent model plans and assertions; text-based plan representation for cliché, functional representation language |
| **Bradman** | pl + cbt | No | No | syntax tree |
| **Adil** | pl + pe + cbt | Yes | Yes | language independent model annotated flow graph |

**Table 2** : A comparison of automated debuggers as tutoring systems (continuation)

\*: ad : application domain, cbt : common bug types, pe : programming expertise, pl : programming language
V : Verification,  C : Checking,  F : Filter,  NA : Not Available

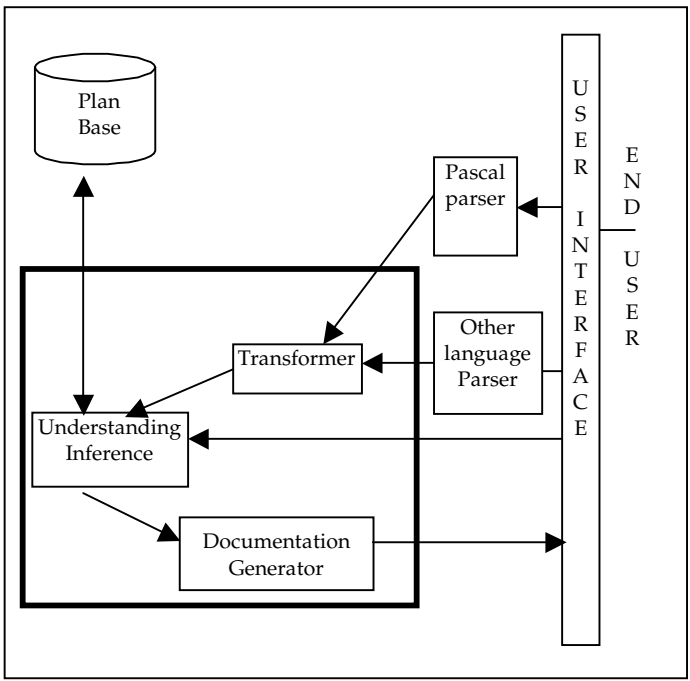 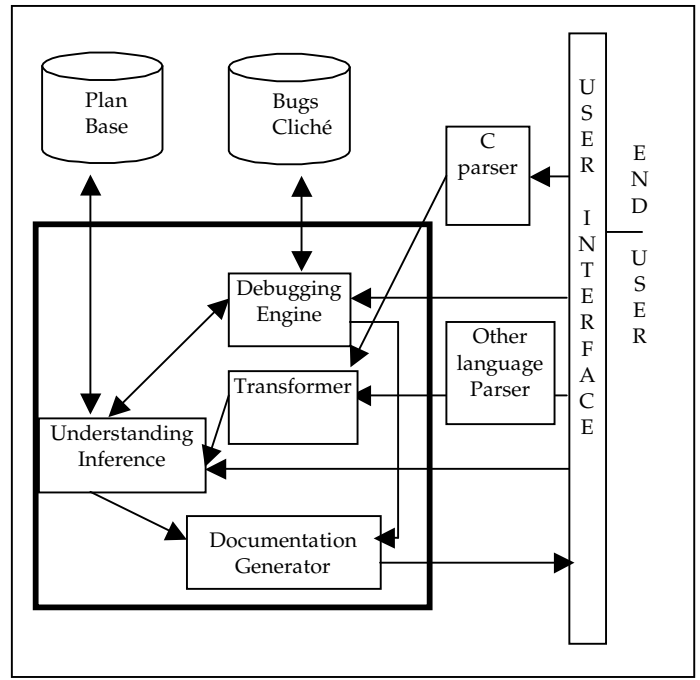

**FIGURE 1**: Conceiver system  **FIGURE 2:** Adil system

We will locate and explain the bugs found. Explaining the bug is a central problem to be addressed by Adil as part of our program understanding framework. Adil will function as an intelligent debugging assistant to guide the students and practically impart the debugging skills through the debugging sessions. We will however refrain from correcting their code. Not only it is contrary to what we have set out above, but also because automatic correction may in fact propagate the bugs further as is evident in the Hamming code problem.

A very important characteristic of Adil is its programming language independent debugging system. This is achieved via the transformed internal representation of the source code as mentioned below. Another important characteristic of Adil is it will automatically construct the knowledge base.

Adil (Fig. 2) will use the constraint satisfaction approach. Initially, two knowledge bases will be constructed; one each for the plan base and the bugs cliché. The C parser will convert and manipulate the source code into an intermediate and transformed parse tree, which in turn, will be converted into an annotated flow graph. The annotated flow graph with data and control flow information will be fed into the inference engine and checked with the library of plans in the knowledge base for the unification of the plans and the code. The understanding inference will try to understand the program and pass the necessary information to the debugging engine which will localize and pinpoint the bugs, if they exist, in the code as output of the debugging system. Lastly, the documentation generator will explain the bug to the user.

### 6. Conclusion

We have proposed a new knowledge-based automated debugger to be used as a user-friendly tool by the students to self-debug their own programs. Stereotyped code (cliché) and bugs cliché will be stored as library of plans in the knowledge-base. Given a program and its specification, Adil will be able to understand a semantic error free program as well as locate, pinpoint and explain logical errors of syntax error-free programs.

Among the major contribution of Adil is its usability, acts as an intelligent debugging assistant for the students, language independent debugging capability, a concise and unambiguous plan formalism, an automatic plan parser tool and a plan base manager tool. Adil can also semi-automatically acquire knowledge (plans), operates as an integrated development environment by containing the necessary supporting tools to highly facilitate the recognition and debugging.

# References

[AL80]   A. Adam and J.-P. Laurent. Laura, a system to debug student programs. *Artificial Intelligence*, 15(1,2):75-122, November 1980.

[Alle91]  D. Allemang. The Use of Functional Models in the Domain of Automatic Debugging. *IEEE Expert*, Dec. 1991.

[Al-Om99] Hani Moh'd Ahmad Al-Omari. CONCEIVER: A Program Understanding System, *PhD Thesis*, Universiti Kebangsaan Malaysia, Bangi, 1999.

[Duca93] Mireille Ducassé. A pragmatic survey of automated debugging. P. Fritzson, editor, in *Proceedings of the first International Workshop on Automated and Algorithmic Debugging*, Linkoping, Sweden, May 1993, Lecture Notes in Computer Science 749, Springer-Verlag, pp 1-15.

[HN88]  M.T. Harandi and J.Q. Ning. PAT: A Knowledge-Based Program Analysis Tool. *Proc. Conf. Software Maintenance*, Phoenix, Ariz., pp. 312-318, 1988.

[JS85] W.L. Johnson and E. Soloway. Proust: Knowledge-based program understanding. *IEEE Transactions on Software Engineering*, SE-11(3):267-275, March 1985.

[Looi88] C-K. Looi. Analyzing novices' programs in a Prolog intelligent teaching system. In *Proceedings of the European Conference on Artificial Intelligence*, pp. 314-319, Munich, August 1988.

[Luke80]  F.J. Lukey. Understanding and debugging programs. *International Journal Man-Machine Studies*, 12(2):189-202, February 1980.

[FR99]  M. Francel and S. Rugaber. The Relationship of Slicing and Debugging to Program Understanding. In *Proceedings of the 7th International Workshop on Program Comprehension*, Pittsburgh, May 1999.

[KN94]  W. Kozaczynski, J. Ning. Automated program understanding by concept recognition, *Automated Software Engineering*, V 1, 1994, pp 61-78.

[Murr88]  W.R. Murray. Automatic Program Debugging for Intelligent Tutoring Systems. *Research Notes in Artificial Intelligence*. Pitman, London, 1988.

[QYW98]  Alex Quilici, Qiang Yang, and Steven Woods. Applying Plan Recognition Algorithms To Program Understanding. *Journal of Automated Software Engineering*, 5(3):347-372, 1998.

[RW90]  Charles Rich, Linda M. Wills. Recognizing a Program's Design: A Graph-Parsing Approach. *IEEE Software*, 7(1):82-89, January 1990.

[Ruth76]  G.R. Ruth. Intelligent Program Analysis. *Artificial Intelligence*, 7(1):65-85, 1976.

[Shah91]  N. Shahmehri. Generalized Algorithmic Debugging. *Phd Thesis*, Linkoping University, 1991.

[SW95]  Philip A Smith, Geoffrey I Webb Transparency Debugging with Explanations for Novice Programmers. in *Proceedings of the 2nd International Workshop on Automated and Algorithmic Debugging*, Saint Malo, France, 1995.

[Wert87]  H. Wertz. Automatic Correction and Improvement of Programs. *Artificial Intelligence*, J. Campbell ed. Ellis Horwood, England, 1987.

[Will92]  Linda M. Wills. Automated Program Recognition by Graph Parsing, MIT-AI-TR 1358 , MIT Artificial Intelligence Laboratory, *Doctoral Dissertation*, July, 1992.

[WY96]  Steven Woods and Qiang Yang. Approaching The Program Understanding Problem: Analysis and A Heuristic Solution, in *Proceedings of the 18th International Conference on Software Engineering*, Berlin, Germany, 1996.